\documentclass[final,1p,times]{elsarticle}
\usepackage{tikz, graphicx}
\usepackage{textpos}
\usepackage{tikz}
\usepackage{ctable}
\usepackage{graphics}
\usetikzlibrary{quantikz}
\usepackage{lineno,hyperref}
\modulolinenumbers[5]
\newdefinition{rmk}{Definitation}
\usepackage{lscape}
\usetikzlibrary{quantikz}
\usepackage{amsthm}
\usepackage{caption}
\usepackage{subcaption}

\begin{document}
	
	\begin{frontmatter}
		
		\title{Quantum multi-secret sharing scheme with access structures and cheat identification}

		\author[mymainaddress]{Deepa Rathi}
		\ead{km.deepa@mt.iitr.ac.in}
		 
		 \author[mymainaddress,secondadress]{Sanjeev Kumar\corref{mycorrespondingauthor}}\cortext[mycorrespondingauthor]{Corresponding author}
		  \ead{sanjeev.kumar@ma.iitr.ac.in}

         \address[mymainaddress]{Department of Mathematics, Indian Institute of Technology Roorkee, Roorkee, India}
        \address[secondadress]{Mehta Family School of Data Science and Artificial Intelligence, Indian Institute of Technology Roorkee, Roorkee, India}
          
		
  \begin{abstract}
  This work proposes a $d$-dimensional quantum multi-secret sharing scheme with a cheat detection mechanism. The dealer creates multiple secrets and distributes the shares of these secrets using multi-access structures and a monotone span program. The dealer detects the cheating of each participant using the Black box's cheat detection mechanism. To detect the participants' deceit, the dealer distributes secret shares' shadows derived from a randomly invertible matrix $X$ to the participants, stored in the black box. The Black box identifies the participant's deceitful behavior during the secret recovery phase. Only honest participants authenticated by the Black box acquire their secret shares to recover the multiple secrets. After the Black box cheating verification, the participants reconstruct the secrets by utilizing the unitary operations and quantum Fourier transform. The proposed protocol is reliable in preventing attacks from eavesdroppers and participants. The scheme's efficiency is demonstrated in different noise environments: dit-flip noise, $d$-phase-flip noise, and amplitude-damping noise, indicating its robustness in practical scenarios. The proposed protocol provides greater versatility, security, and practicality.
		\end{abstract}
		\begin{keyword}
           Black box \sep Cheat identification \sep Multi access structure \sep Noise environments \sep Quantum Fourier transform 
		\end{keyword}
		
	\end{frontmatter}

\section{Introduction}
In today's world, where hackers continually target secret data, secret sharing is a crucial cryptographic technique that assures the security and confidentiality of secret information. Shamir\cite{shamir1979share} and Blakley\cite{blakley1979safeguarding} devised the first threshold secret sharing technique separately, employing Lagrange's interpolation and projective geometry theories, respectively. However, classical secret sharing schemes rely on mathematical assumptions and computational complexity, which cannot provide secure demonstrative communication. C.H. Bennett and G. Brassard devised the famed BB84\cite{brassard1984quantum} protocol to solve the limits of classical cryptography techniques. This protocol is considered to be the beginning of quantum cryptography. The absolute security of quantum cryptography relies on the fundamental characteristics of quantum mechanics, including the no-cloning theorem, the Heisenberg uncertainty principle, and the inability to distinguish non-orthogonal quantum states.
\\
Quantum secret sharing (QSS) is a significant field of study in quantum cryptography. This cryptographic technique provides enhanced security and several advantages over classical schemes. QSS research is often separated into two groups based on shared secrets: Quantum state sharing (QSTS) for sharing unknown quantum states and QSS for sharing classical information. In 1999, Hillery et al.\cite{hillery1999quantum} introduced the founding work of QSS by utilizing the entangled three-qubit and four-qubit GHZ states. At the same time, $(t,m)$-threshold QSS protocols have been introduced \cite{cleve1999share,karlsson1999quantum}, in which at least $t$ out of $m$ participants are required to retrieve the secret. Gottesman\cite{gottesman2000theory} proved that the no-cloning theorem and monotonicity constraints are sufficient for the existence of QSS schemes with access structures. Subsequently, Xiao et al.\cite{xiao2004efficient} extended the QSS scheme\cite{hillery1999quantum} by implementing the quantum key distribution techniques: favored measuring basis and measuring basis encrypted. Deng et al.\cite{deng2005multiparty} envisioned a QSTS scheme to share an arbitrarily two-qubit state by utilizing Einstein-Podolsky-Rosen (EPR) pairs. Henceforth, QSS protocols of $2$-dimensional quantum system have been extensively studied in  Ref.\cite{hsu2003quantum,markham2008graph,jia2012dynamic,cao2018verifiable,musanna2020novel,musanna2022quantum}.
\\
Most QSS protocols discussed before are built for $2$-dimensional quantum systems (qubits). However, with advancements in quantum technology, developing QSS protocols in high dimensional quantum systems (qudits) is becoming more significant than qubits due to their higher information capacity and improved security. Therefore, several QSS protocols are presented in high-dimensional quantum systems. Yu et al.\cite{yu2008quantum} developed a $d$-dimensional QSS protocol with mutually unbiased and biased bases by generalizing the two-qubit QSS scheme\cite{hillery1999quantum}. Tavakoli et al.\cite{tavakoli2015secret} introduced a multiparty QSS scheme by utilizing a sequential communication of a single quantum system with $d$-dimensional. Subsequently, Chen et al.\cite{chen2018cryptanalysis} demonstrated that the protocol reported in \cite{tavakoli2015secret} needs to be more secure and efficient. They evaluated the security vulnerabilities and enhanced the efficiency, growing it from $1/d$ to $1$. Using the quantum Fourier transform (QFT), and generalized unitary operators, Song et al.\cite{song2017t} presented the $(t,m)$-threshold QSS scheme. The secret is retrieved by the reconstructor employing the inverse quantum Fourier transform (IQFT) without relying on any information from the remaining participants. However, Kao et al.\cite{kao2018comment} discovered that in the scheme \cite{song2017t}, the reconstructor cannot retrieve the secret without the help of other participants. Later, Sutradhar and Om\cite{sutradhar2021enhanced} overcome this problem by introducing an enhanced $(t,m)$ threshold QSS protocol. Qin et al.\cite{qin2018multi} presented a QSTS scheme by utilizing the QFT.          
\\
QSS systems are categorized into two groups based on the number of participants in authorized sets: threshold and general. The current QSS methods are mainly $(t,m)$-threshold\cite{qin2016verifiable,song2017t,lu2018verifiable,song2019verifiable,sutradhar2021enhanced,bai2021verifiable,rathi2023d}, allowing any subset of $t$ participants or more to retrieve the secret, whereas subsets with fewer than $t$ participants are unable to retrieve the secret. In practical situations, the composition of authorized subsets may not depend on $t$, leading to the proposal of general QSS techniques which utilize access structures to determine authorized subsets \cite{wang2014quantum,qin2016d,mashhadi2019general,li2021general,bai2022quantum,mashhadi2022verifiable,wu2022two}. The access structure describes participant subsets that can retrieve the secret, while the adversary structure refers to participant subsets that cannot get any information of the secret. Given that most QSS schemes only consider ideal noise-free quantum channels, i.e., without considering the impact of channel noise on the QSS schemes in real quantum communication. Nevertheless, in real quantum communication, the quantum states must engage via the surrounding environment. Which introduces influences from channel noise and disrupts quantum resource entanglement. Thus, studying the effect of QSS protocols in noisy environments is essential. Some QSS schemes with a $2$-dimensional quantum system in noisy environments have been reported in Ref.\cite{wang2015secret,khakbiz2019sequential,bai2020improving,huang2021quantum,hu2022conclusive}.
\\
Furthermore, in most of the QSS, as mentioned earlier protocols, the dealer and participants can recognize if there is cheating but cannot identify the culprit. Yan et al.\cite{yan2022cheating} introduced a threshold QSS protocol to identify cheaters using a voting mechanism. However, this technique is not analyzed in noisy environments, limiting its feasibility and application versatility. Li et al.\cite{li2022new} introduced a cheating-detectable classical secret sharing method to detect and remove cheaters using asymmetric bivariate polynomial and Black box methodology. In their scheme, the $m$ participants are divided into $t$ disjoint sets, and one trusted dealer is assigned to each group. Nevertheless, this method may not be appropriate for some practical situations. Therefore, We are considering combining the Black box deception algorithm with QSS to make the scheme general and unconditionally secure.\\
We study a cheating identifiable quantum multi-secret sharing (QMSS) scheme with general access structures. In QMSS, multiple secrets are distributed to the participants simultaneously. The dealer assigns $n$ secrets to the participants according to $n$ distinct access structures employing a monotone span program (MSP) and linear multi-secret sharing (LMSS). During the recovery phase, the participants' cheating behavior was identified utilizing the Black box's deception verification mechanism. After the cheating verification, the participants directly exchange their secret shares through the Black box and then regenerate the secrets. The participants implement the generalized Pauli operator and QFT to retrieve the secret, and a hash function is utilized to validate the authenticity of secrets. Moreover, we evaluate the effectiveness of the proposed scheme in three kinds of noise models: dit-flip, d-phase-flip, and amplitude-damping observed in real-world scenarios. The proposed scheme distinguishes itself from existing QSS methods in the following ways:
\begin{enumerate}
 \item The scheme is feasible to share multiple secrets simultaneously based on multi-access structures.
\item  Each participant's deception is identified by a Black box.
 \item The proposed scheme is independent of trustworthy third parties due to the cheating verification mechanism.
 \item It can withstand participant attacks, including forgery and collusion attacks.
\item The influence of noisy environments on the proposed QMSS is demonstrated through fidelity. 
\end{enumerate}

 \section{Preliminaries}	
\subsection{Unitary operators}
\begin{rmk}
The generalized Pauli operator for a qudit system with dimension $d$ is specified as
\begin{equation*}
    U_{a,b} = \sum_{z=0}^{d-1} \omega^{b z} \ket{z+a} \bra{z},
\end{equation*}
where $\omega = e^{\frac{2\pi i}{d}}$, and $ a, b \in \{0,1,...,d-1\}$.
\end{rmk}
\begin{rmk}
The quantum Fourier transform $ \mathcal{F}$ executed on a quantum state $\ket{x}$ of $d$-dimensional is written as
    \begin{equation*}
        \mathcal{F} \ket{x} = \frac{1}{\sqrt{d}}\sum_{z=0}^{d-1} \omega^{x z} \ket{z}, \text{where} ~~ \omega = e^{\frac{2\pi i}{d}}.
    \end{equation*}
    The inverse quantum Fourier transform $\mathcal{F}^{-1}$ applied on a qudit state $\ket{z}$ is represented by
     \begin{equation*}
        \mathcal{F}^{-1} \ket{z} = \frac{1}{\sqrt{d}}\sum_{x=0}^{d-1} \omega^{-z x } \ket{x}, \text{where} ~~ \omega = e^{\frac{2\pi i}{d}}. 
    \end{equation*}
\end{rmk}
\begin{rmk}
  The quantum SUM gate for two qudits $\ket{\alpha}$ and $\ket{\beta}$ is written as
   \begin{equation*}
       SUM(\ket{\alpha},\ket{\beta}) = (\ket{\alpha},\ket{\alpha+\beta}). 
   \end{equation*}
   In this context, $\ket{\alpha}$ represents the control particle, $\ket{\beta}$ represents the target particle, and $``+"$ indicates the addition modulo $d$.
\end{rmk}

\subsection{Access structure}	
\begin{rmk}
    Let $\Omega = \{P_{1},P_{2},...,P_{m}\}$, is a collection of participants and $\Gamma$ be a subset of $2^{\Omega}$. A $\Gamma \subseteq 2^{\Omega}$ access structure can be considered as a set of authorized participants if satisfies the conditions: $B\in \Gamma$ when $A\in \Gamma,~ A\subseteq B \subseteq \Omega$. The adversary structure $\Delta$ refers to the collection of unauthorized sets, i.e., $\Delta = \Gamma^{c}$.
\end{rmk}

\begin{rmk}
For a secret $s_{i}$, the access structure $\Gamma_{i} \subseteq 2^{\Omega}$ is a family of sets of authorized participants to get the secret $s_{i}$. A multi access structure $\Gamma = (\Gamma_{1},\Gamma_{2},...,\Gamma_{n})$ consisting of $n$ sets is used for $n$ secrets $(s_{1},s_{2},...,s_{n})$.
\end{rmk}

\begin{rmk}
  A monotone span program (MSP) is represented by $(Z_{d},M,\psi, \zeta_{i})$, where $Z_{d}$ be a finite field ($d$ is a prime), $M$ is a matrix of $m \times l$ order over $Z_{d}$,  $\psi : \{1,2,...,m\} \rightarrow \Omega$ is a surjection map used to assign the rows of $M$ to each participant, and $\zeta_{i} = (0,...,0,1,0,...,0)^{T} \in Z_{d}^{l}$ (where $1$ is the $i$th element) is the target vector.   
\end{rmk}

\begin{rmk}
For multi-access structure $\Gamma = (\Gamma_{1},\Gamma_{2},...,\Gamma_{n})$, if $(Z_{d},M,\psi, \zeta_{i})$, $i=1,2,...,n$, satisfies the following conditions then it is referred to as a monotone span program (MSP). 
\begin{enumerate}
    \item[(1)] For any $A \in \Gamma_{i}$, there exists a vector $\lambda_{iA}$ such that $M_{A}^{T}\lambda_{iA} = \zeta_{i}$.
    \item[(2)] For any $A \in \Delta_{i}$, there exists a vector $\kappa = (\kappa_{1},...,1,...,\kappa_{l-1})^{T}\in Z_{d}^{l}$ such that $M_{A}\kappa = 0 \in  Z_{d}^{l}$ with $1$ is the $i$th element.
    \end{enumerate}
   In this context, $M_{A}$ represents the rows $k$ of $M$ where $\psi(k) \in A$, and $T$ signifies the transpose.
\end{rmk}

\subsection{Linear multi-secret sharing (LMSS)}
Linear multi-secret sharing (LMSS) is considered one of the most efficient methodologies in the field of general secret sharing. The LMSS could be utilized for access control techniques for large data sets with minimal additional cost. Following the MSP $(Z_{d},M,\psi, \zeta_{i})$, we examine the formulation of an LMSS about multi-access structure  $\Gamma = (\Gamma_{1},\Gamma_{2},...,\Gamma_{n})$. The dealer $D$ wants to distribute the $n$ secrets $s_{1},s_{2},...,s_{n}\in Z_{d},$ to $m$ participants using the multi access structure $\Gamma = (\Gamma_{1},\Gamma_{2},...,\Gamma_{n})$. The dealer $D$ examined a MSP  $(Z_{d},M,\psi, \zeta_{i})$.
\begin{enumerate}
\item[(1)] \textbf{Distribution phase:} The dealer $D$ calculates the shares of the participant by selecting a random vector $\rho = (s_{1},...,s_{n},\rho_{n+1},...,\rho_{l})\in Z_{d}^{l}$. Then, $D$ calculates $sh = M\rho = (sh_{1},sh_{2},...,sh_{m})^{T}$ and distribute the share $sh_{k}$ among the participant $\psi(k)$ via secure quantum channel.  
    
\item[(2)] \textbf{Reconstruction phase:} Consider that $A\in \Gamma_{i}$ and $sh_{A}$ represents the elements of $sh$ that have indices in the set $A$. The participants of the set $A$ regenerate the $i$th secret $s_{i}$ as:
 \begin{equation*}
        sh_{A}\lambda_{iA} = (M_{A}\rho)^{T} \lambda_{iA} = \rho^{T}(M_{A}^{T}\lambda_{iA}) = \rho^{T} \zeta_{i} = s_{i}.
    \end{equation*}
\end{enumerate}

\noindent\textbf{Remarks:}
\begin{enumerate}
 \item[1] For any set $A \subseteq \Omega,$ if $A \not\subset \Gamma,$ then $A \subseteq \Gamma^{c} = \Delta $.
    \item[2] An unauthorized subset of $\Delta$ cannot acquire all the secret shares, whereas an authorized subset of $\Gamma$ obtains all secret shares. 
    \item[3] If $\omega = e^{\frac{2 \pi i}{d}}$, then 
\begin{equation*}
        \sum_{y=0}^{d-1} \omega^{xy} =  \begin{cases}
        d, & x \overset{d}{\equiv}0; \\
        0, & x \overset{d}{\not\equiv}0.
        \end{cases} 
  \end{equation*}

\end{enumerate}
   
\subsection{Black box mechanism for cheat-identification}
The term ``Black box"\cite{li2022new} means that a device or product's internal structure or principles are not significant to the user. Thus, the user is only interested in the device's functionality and how to operate it.\\
In our protocol, the Black box is required to execute the following functions:
\begin{enumerate}
  \item The dealer $D$ develops a diagonal matrix $\Sigma$ of $2m$-order of secret shares $sh_{k}$ and computes the matrix $X=Y^{-1} \Sigma Y$. $D$ determines two independently eigenvectors $(y_{k1},y_{k2})$ corresponding to the eigenvalues of $X$, and $(y_{k1},y_{k2})$ are utilized as the shadows of secret shares. These shadows $(y_{k1},y_{k2})$ are transmitted to the participants. Then, these $sh_{k}$ and matrix $X=Y^{-1} \Sigma Y$ are kept in the Black box. 
  \item In the reconstruction phase, the Black box validates the secret shares' shadow given by the participants. Therefore, the following two factors are used to verify the existence of cheaters:
 \begin{itemize}
        \item[$\bullet$] $y_{k1}$ and $y_{k2}$ are linearly independent.
         \item[$\bullet$]  $sh_{k} = sh_{k1} = sh_{k2}$, where $sh_{k1}$ and $sh_{k2}$ can be evaluated by solving the equations $X y_{k1} = sh_{k1} y_{k1}$ and $X y_{k2} = sh_{k2} y_{k2}$, respectively.
 \end{itemize}
 \item After the cheating verification of participants, the Black box transmits the secret shares $sh_{k}$ to the participants $\psi(k)$ through a secure quantum channel.      
\end{enumerate}
\subsection{Noise models}
The operator sum representation efficiently depicts the interaction between a quantum state and its surrounding environment. Using Kraus operators\cite{fonseca2019high}, the noise model for $d$-dimensional quantum states may be characterized by an entirely positive trace-preserving map $\epsilon$.
\begin{equation*}
    \rho' = \epsilon(\rho) = \sum_{m',n'} E_{m',n'} \rho E_{m',n'}^{\dagger}
\end{equation*}
where $E_{m',n'}^{\dagger}$ denotes the conjugate transpose of $E_{m',n'}$, $\rho$ and $\rho'$, are the density matrices of the input quantum state and corresponding output quantum state,  respectively. The Kraus operators $E_{m',n'}$ are associated to the Weyl operators $\hat{U}_{m',n'}$\cite{bertlmann2008bloch} described as:
\begin{equation*}
    \hat{U}_{m',n'} = \sum_{z=0}^{d-1} \omega^{m'z} \ket{z} \bra{z+n'} 
\end{equation*}
where $``+"$ means addition modulo $d$.\\
The widely recognized noise models\cite{fonseca2019high} in quantum channels, dit-flip, d-phase-flip, and amplitude damping, represented as:
\begin{enumerate}
    \item \textbf{Dit-flip noise:} This noise involves disturbances that convert $\ket{z}$ with probability $\mu$, either to the state $\ket{z+ 1}$, $\ket{z+ 2},..., \text{or} \ket{z+ d-1}$, whereas preserving it unaltered with the probability $1-\mu$. The associated Kraus operators are represented  as:
    \begin{equation*}
        E_{0,0}=\sqrt{1-\mu}\hat{U}_{0,0},~ E_{0,1}=\sqrt{\frac{\mu}{d-1}}\hat{U}_{0,1},...,E_{0,d-1}=\sqrt{\frac{\mu}{d-1}}\hat{U}_{0,d-1}
    \end{equation*}

     \item \textbf{d-phase-flip noise:} This noise refers to the phenomenon where quantum information is lost without energy dissipation. In this noise, the state $\ket{z}$ is susceptible to a phase transformation with a probability of $\mu$, resulting in one of the $d-1$ phases: $\omega\ket{z}$, $\omega^2\ket{z}$, ..., or $\omega^{d-1}\ket{z}$. The Kraus operators are shown as:
 \begin{equation*}
        E_{0,0}=\sqrt{1-\mu}\hat{U}_{0,0},~ E_{1,0}=\sqrt{\frac{\mu}{d-1}}\hat{U}_{1,0},...,E_{d-1,0}=\sqrt{\frac{\mu}{d-1}}\hat{U}_{d-1,0}
    \end{equation*}

    \item \textbf{Amplitude-damping noise:} The consequences of energy dispersion in a quantum system caused by energy loss are referred to as amplitude-damping noise. This noise will change the basis state $\ket{z}$ to the state $\ket{0}$ with a probability of $\mu$ excluding the state $\ket{0}$, and leave it unchanged with a probability of $1- \mu$. The associated Kraus operators are represented as:
  \begin{equation*}
    E_{0}= \ket{0}\bra{0} + \sqrt{1-\mu} \sum_{z=1}^{d-1} \ket{z}\bra{z}, ~ E_{z}= \sqrt{\mu} \ket{0}\bra{z},~\text{with}~ z=1,2,...,d-1.
\end{equation*}
\end{enumerate}

The density matrix for $m$-qudit state through Kraus operators is described as:
\begin{equation*}
    \rho' = \epsilon(\rho) = \sum_{r_{1},r_{2},...,r_{n}} (E_{r_{1}}\otimes E_{r_{2}}\otimes...\otimes E_{r_{n}})\rho (E_{r_{1}}\otimes E_{r_{2}}\otimes...\otimes E_{r_{n}})^{\dagger}.
\end{equation*}
Where $E_{r_{z}}$ represents the $z$th qudit influenced by the channel noise.
\\
The influence of noise on the quantum state is visualized by determining the fidelity between the initial quantum state, say $\ket{\phi}$, and the output density matrix $\rho_{out}$. Fidelity quantifies the similarity between two quantum states and is a mathematical measure for assessing their degree of closeness. Fidelity is defined by:
\begin{equation*}
    F= \langle \phi | \rho_{out} | \phi \rangle
\end{equation*}
If $F=1$, no noise exists in the quantum channel. However, $F=0$ indicates that all information has been lost. Thus, $0\leq F\leq 1$.

\section{Proposed QMSS scheme}
\begin{figure*}[htbp!]
    \centering
    \includegraphics[width=13cm,height=6cm]{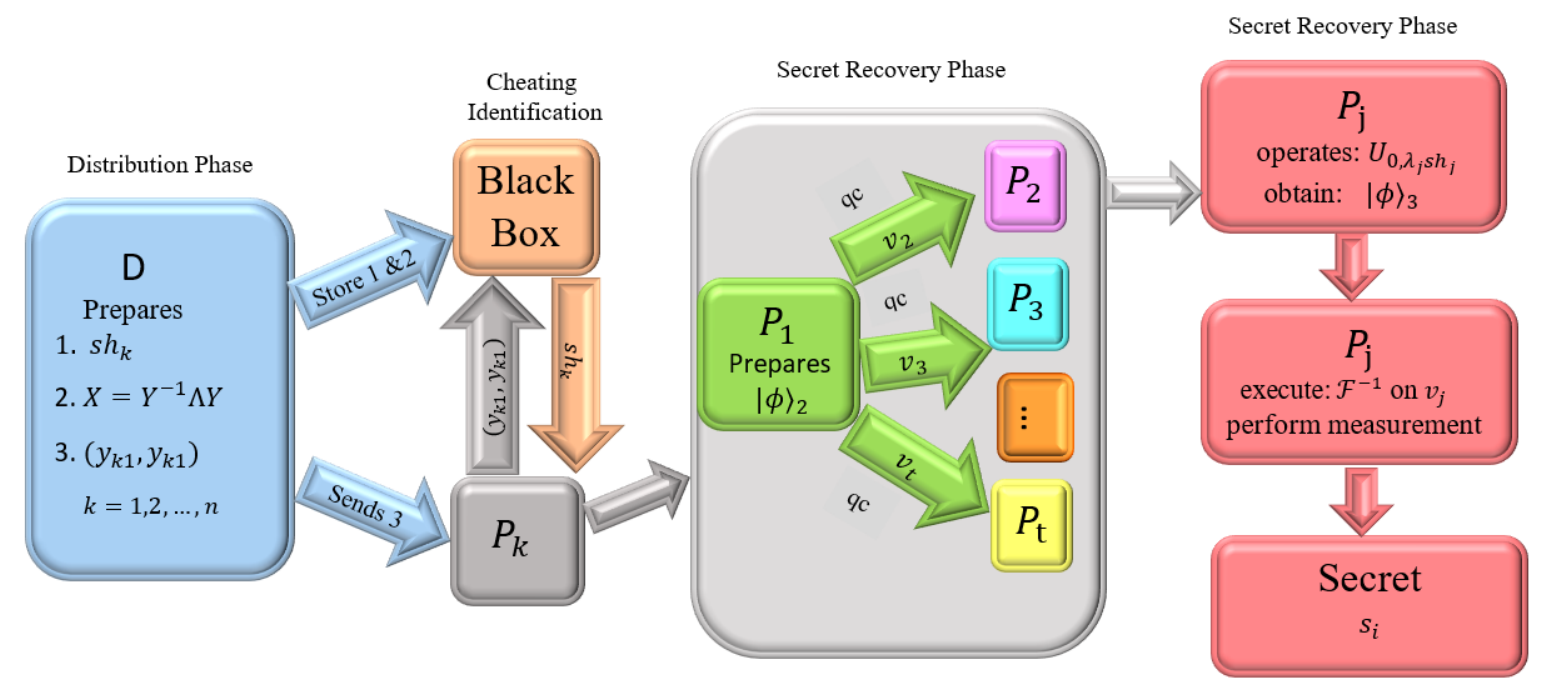}
    \caption{QMSS scheme with the authorized set $A$ for secret $s_{i}$ ($j=1,2,...,t$ and qc = quantum channel).}
    \label{QMSS}
\end{figure*}

The proposed cheating-identifiable quantum multi-secret sharing (QMSS) technique comprises a dealer $D$, $m$ participants $\{P_{1},P_{2},...,P_{m}\}$ and a Black box. Assume that the dealer $D$ intends to allocate $n$ secrets $(s_{1},s_{2},...,s_{n})$ to $m$ participants $\{P_{1},P_{2},...,P_{m}\}$, based on the multi access structures $\Gamma = (\Gamma_{1},\Gamma_{2},...,\Gamma_{n})$. Additionally, $d$ is a prime number, $h$ represents a hash function, and $(Z_{d},M,\psi, \zeta_{i})$ denotes a monotone span program (MSP) for $\Gamma$. The graphical representation of the QMSS scheme is depicted in Fig.\ref{QMSS}.


\subsection{Distribution phase}
The dealer $D$ executes the following actions.
\begin{enumerate}
    \item $D$ select a random vector $ \rho = (s_{1},...,s_{n},\rho_{n+1},...,\rho_{l})^{T}\in Z_{d}^{l}$
    
    \item Compute $sh = M_{m\times l} \rho = (sh_{1},sh_{2},...,sh_{m})^{T}$.

    \item $D$ creates a diagonal matrix $\Sigma$ of order $2m$ with diagonal elements $sh_{k},~k = 1,2,,...,m$ as
    \begin{equation}
      \Sigma =   \text{diag}\Big\{sh_{1},sh_{1},sh_{2},sh_{2},...,sh_{m},sh_{m}\Big\}.
    \end{equation}
  Now, the dealer $D$ randomly develops a $2m$-order invertible matrix $Y$ over $Z_{d}$ and compute $X = Y^{-1} \Sigma Y$. It is known that the eigenvalues of matrices $X$ and $\Sigma$ are identical due to their similarity. There are two linearly independent (LI) eigenvectors $(y_{k1},y_{k2})$ correspond to the eigenvalues $sh_{k}~(k=1,2,...,m)$. Therefore, each eigenvalue must have at least two LI eigenvectors. The dealer $D$ occurs the linearly independent eigenvectors $(y_{k1},y_{k2})$ corresponding to the eigenvalue $sh_{k}$ of participant $P_{k}$ as secret shares' shadows. Subsequently, $D$ transmits each pair of secret shares' shadows $(y_{k1},y_{k2})$ to the participant $\psi(k),~\psi(k) \in \Gamma_{i} $ through secure quantum channel. For simplicity, we assume that $\psi(k) = P_{k}$ for $1\leq k \leq m.$\\
  The secret shares (eigenvalues of matrix $\Sigma$) $sh_{1},sh_{2},...,sh_{m}$ and the matrix $X = Y^{-1} \Sigma Y$ are stored in the Black box. 
 \item Using the public hash function $h$, $D$ calculates and publishes the hash values $H_{i} = h(s_{i})$, $i=1,2,...,n$.
 
\end{enumerate}

\subsection{Reconstruction phase}
Assume that the participants of an authorized set $A\in \Gamma_{i}$ are required to retrieve the secret $s_{i}$. To simplify the explanation, we assume that $A = \{P_{1},P_{2},\ldots,P_{t}\}$.  The cheating of the participants was detected through a Black box mechanism that relies on the matrix $X$. Participants verified as honest through the Black box may acquire the secret shares and then successfully reconstruct the secret $s_{i}$.

\subsubsection{Cheating identification phase}
\begin{enumerate}
    \item Consider that the participants $P_{k},~ k=1,2,...,t,$ provide the shadows $(y_{k1},y_{k2})$ and recover the secret $s_{i}$. The procedure for reconstructing the secret can be executed if the following conditions are met:
    \begin{itemize}
         \item[$\bullet$] $y_{k1}$ and $y_{k2}$ are linearly independent
          \item[$\bullet$] $sh_{k} = sh_{k1} = sh_{k2}$
    \end{itemize}
otherwise, continue with step $2$.
   \item If participant $P_{k}$ is detected as dishonest in the previous step, he will be eliminated. The procedure of secret recovery will be terminated if the set of participants without a cheater is not a subset of $\Gamma_{i}$.
 \item After the cheating verification of all participants, the Black box transmits the secret shares $sh_{k}~(k=1,2,...,t)$ to the participants $P_{k}$ of the authorized set $A\in \Gamma_{i}$ through a secure quantum channel.
\end{enumerate}

\subsubsection{Secret recovery phase}
Assume that after obtaining the secret's shares $sh_{k}$, the participants $P_{k}$ of a set $A\in\Gamma_{i}$ want to recover the secret $s_{i}$. To simplify the explanation, we assume that $A=\{P_{1},P_{2}...,P_{t}\}$ is the qualifying subset of participants, and $P_{1}$ is a reconstructor. The reconstruction process proceeds as follows:
\begin{enumerate}
    \item The participant $P_{1}$ generates $t$ single qudits $\ket{0}_{1},\ket{0}_{2},...,\ket{0}_{t}$.

    \item $P_{1}$ operates the QFT $\mathcal{F}$ on the first particle $\ket{0}_{1}$ and get the state $\ket{\phi_{1}}$ as
    \begin{align}\nonumber
 \ket{\phi_{1}} & = (\mathcal{F} \ket{0}_{1})\ket{0}_{2},...,\ket{0}_{t} \\ &
         = \Big(\frac{1}{\sqrt{d}} \sum_{v=0}^{d-1} \ket{v}_{1}\Big)\ket{0}_{2},...,\ket{0}_{t}.
   \end{align}

   \item $P_{1}$ applies the $t-1$ quantum SUM operations on the particles $\ket{0}_{j},~j = 2,3,..,t$ with $(\mathcal{F} \ket{0}_{1})$ as the control qudit and $\ket{0}_{j},(j = 2,3,..,t)$ as the target qudits. The generated entangled quantum state $\ket{\phi_{2}}$ is 
   \begin{equation}
       \ket{\phi_{2}} = \frac{1}{\sqrt{d}} \sum_{v=0}^{d-1} \ket{v}_{1}\ket{v}_{2}...\ket{v}_{t}.
   \end{equation}

   \item $P_{1}$ distributes the particle $\ket{v}_{j},~j= 2,3,...,t$ to the participants $P_{j}$ respectively, via the secure quantum channels.

   \item Every participant $P_{j}~(j=1,2,...,t)$ operates the Pauli operators $U_{0,\lambda_{j}sh_{j}}$ on their respective particles $\ket{v}_{j~~}$, to obtain the quantum state $\ket{\phi_{3}}$ as:
   \begin{align} \nonumber
      \ket{\phi_{3}} &= U_{0,\lambda_{1}sh_{1}} \otimes U_{0,\lambda_{2}sh_{2}}\otimes...\otimes U_{0,\lambda_{t}sh_{t}} \ket{\phi_{2}} \\ \nonumber &
             = \frac{1}{\sqrt{d}} \sum_{v=0}^{d-1} \omega^{\lambda_{1}sh_{1}v}\ket{v}_{1}\omega^{\lambda_{2}sh_{2}v}\ket{v}_{2}...\omega^{\lambda_{t}sh_{t}v}\ket{v}_{t} \\  &
             = \frac{1}{\sqrt{d}} \sum_{v=0}^{d-1} \omega^{(\sum_{j=1}^{t}\lambda_{j}sh_{j})v} \ket{v}_{1}\ket{v}_{2}...\ket{v}_{t}.
   \end{align}

\item Each participant $P_{j}$ executes the inverse quantum Fourier transform $\mathcal{F}^{-1}$ on their particle $\ket{v}_{j}$, and then measures the outcomes. After performing measurements on the particles, each participant $P_{j}$ broadcasts his measurement result.

\item The participants $P_{j}~(j=1,2,...,t)$ sum up their measurement outcomes and compute the secret $ \sum_{j=1}^{t}\lambda_{j}sh_{j}~\text{mod} d = s_{i}$.

\item Each participant $P_{j}$ checks the recovered secret by $H_{i} = h(s_{i})$, where $h$ is a hash function. If this test is correct, they can conclude that all participants are trustworthy; otherwise, they ensure that any of the participant is deceitful.

\end{enumerate}

\section{Security analysis}
This section examines the proposed scheme's security against internal and external attackers and demonstrates its resistance to their actions.

\subsection{Intercept resend attack}
Assume that the eavesdropper, Eve, has control over the quantum channel. Then, Eve intercepts the qudits $\ket{v}_{j}$ and measures them on a computational basis to obtain secret information. Additionally, Eve creates and resends the fictitious particle $\ket{v'}_{j}$ to $P_{j}$. After measuring the particle, Eve may obtain the corrected value $v$ with a probability of $1/d$. However, Eve is unable to acquire any information about the secret shadows $(y_{k1},y_{k2})$ and the secret $s_{i}$. Since the transmitted particles $\ket{v}_{j}$ contain no information of secret shadows and secret shares. 

\subsection{Entangle measure attack}
The eavesdropper Eve obtains all of the particles $\ket{v}_{j} (j=2,3,...,t)$, when $P_{1}$ transmits the particles $\ket{v}_{j}$ to participants $P_{j}$. Afterward, Eve proceeds to create an additional particle $\ket{a}$ and entangles it with one of the intercepted particles $\ket{v}_{j}$. Eve applies the SUM operator on the particles $\ket{a}$ and $\ket{v}_{j}$. Thus, the state $\ket{\phi_{2}}$ develops into $\ket{\phi_{2}}'$ as  
\begin{equation}
    \ket{\phi_{2}}' = \frac{1}{\sqrt{d}} \sum_{v=0}^{d-1} \ket{v}_{1}\ket{v}_{2}...\ket{v}_{t}\ket{v+a}. 
\end{equation}
Subsequently, Eve chooses another secret particle $\ket{v}_{r}$ and executes a SUM operator on $\ket{a}$. Consequently, the quantum state $\ket{\phi_{2}}'$ evolves into $\ket{\phi_{2}}''$
\begin{equation}
    \ket{\phi_{2}}'' = \frac{1}{\sqrt{d}} \sum_{v=0}^{d-1} \ket{v}_{1}\ket{v}_{2}...\ket{v}_{t}\ket{v+v+a} = \ket{\phi_{2}} \ket{a}. 
\end{equation}
Eve acquires the initial value $a$ by measuring the ancillary particle $\ket{a}$. Consequently, he concludes that $\ket{v}_{j}$ and $\ket{v}_{r}$ are equivalent. Similarly, he can only conclude that all transmitted particles $\ket{v}_{j}$ are identical. Thus, Eve cannot gain any secret information from the intercepted particles $\ket{v}_{j}$. 

\subsection{Collusion attack}
The secret shares $sh_{k}$ are exclusively held by the dealer $D$, while the participants can only get the shadows $(y_{k1},y_{k2})$ of the secret shares. Therefore, even if they collaborate, the participants cannot recreate the secret shares to retrieve the secret $s_{i}$. Only the participants authenticated by the Black box can acquire all the correct secret shares $sh_{k}$. Hence, the secrets $s_{i}~ (i=1,2,...,n)$ remain secure. Alternatively, in the secret recovery phase, each participant $P_{j}$ measures their particle in the computational basis and publicly announces the outcome of their measurement $\lambda_{j}sh_{j}$. However, this process does not reveal the value shares of participant $P_{j}$ to the other participants.\\
Moreover, assume that participants of an unauthorized set $C \subset A$ conspire to obtain additional secret shares from the shared particles. Their assault will not succeed in the proposed scheme. Since only participant $P_{1}$ distributes the secret particles, $\ket{v}_{j}$ to the rest and $\ket{v}_{j}$ carries no secret information. Additionally, if participants from an unauthorized set $C \subset A$ try to access the secret $s_{i}$, they would be required to get the secret shares held by the other participants in set $A$. However, as explained earlier, they cannot obtain these shares. Therefore,  according to the LMSS, unauthorized participants cannot compute the secret by performing linear operations on their shares.

\subsection{Forgery attack}
Suppose that the participant $P_{k}$ provides fake shadows during the cheating verification process. The secret shares $sh_{k}$ and the matrix $X = Y^{-1} \Sigma Y$ are stored in the Black box. The Black box validates the participant's shadows based on the following two conditions:
\begin{itemize}
        \item[$\bullet$] $y_{k1}$ and $y_{k2}$ are linearly independent.
         \item[$\bullet$]  $sh_{k} = sh_{k1} = sh_{k2}$, $sh_{k1}$ and $sh_{k2}$ can be determined by calculating the equations $X y_{k1} = sh_{k1} y_{k1}$ and $X y_{k2} = sh_{k2} y_{k2}$, respectively.
        \end{itemize} 
Consequently, the eigenvalues must be consistent without cheating when compared to $sh_{k}$ stored in the Black box. If a participant provides fake shadows, these two conditions are not satisfied,  and the participant is identified as a cheater. Hence, the participants cannot forge the secret shares' shadows.\\
During the reconstruction process, every participant provides the secret shares' shadow $(y_{k1},y_{k2})$  rather than the required information $sh_{k}$. Although the attackers obtain $(y_{k1},y_{k2})$, they have to compute $X y_{k1} = sh_{k1} y_{k1}$ and $X y_{k2} = sh_{k2} y_{k2}$  to acquire $ sh_{k1}, sh_{k2}$, respectively. However, only the dealer $D$ and the Black box know about the matrix $X$. Therefore, no one can obtain information of $sh_{k}$ from $(y_{k1},y_{k2})$. Only the participants authenticated by the black box can obtain all the secret shares. Furthermore, since the Black box directly transmits these $sh_k$, the internal attacker cannot fabricate the recovery of the secret.\\
In the secret recovery process, assume that certain malicious participants within the authorized set $A$ utilize a Pauli operator with a counterfeit share. As a result, each participant calculates incorrect values for the secret $s_{i}$ and obtains $h(s_{i}) \neq H_{i}$. Then, they conclude that some participants are dishonest.  


\section{Example}
Let $\Omega = \{P_{1},P_{2},P_{3},P_{4}\}$ represent the set of participants and $\Gamma = (\Gamma_{1}, \Gamma_{2})$ indicates the access structures with $\Gamma_{1} = \{A_{1} = \{P_{1},P_{2},P_{3}\}$, $A_{2} = \{P_{1},P_{2},P_{4}\}, A_{3} = \Omega \}, \Gamma_{2} = \{A = \Omega\}  $ . Assume the dealer $D$  intends to reveal two secret $s_{1}=2$ and $s_{2}=5$ to four participants, using the multi access structure $\Gamma = (\Gamma_{1}, \Gamma_{2})$ and MSP $(Z_{7},M,\psi,\zeta_{1},\zeta_{2})$. The labeling map $\psi(k)=P_{k}$, $ \forall ~k\in\{1,2,3,4\}$, $ \zeta_{1} = (1,0,0,0)^{T}$, $\zeta_{2} = (0,1,0,0)^{T},$ and 
$ M = \begin{bmatrix} 4 & 1 & 1 & 1 \\ 0 & 0 & 1 & 1 \\6 & 3 & 0 & 0 \\0 & 1 & 1 & 1          \end{bmatrix}$.\\
So, $\lambda_{1A_{1}} = (4,3,1)^{T},\lambda_{1A_{2}} = (2,0,5)^{T}, \lambda_{1A_{3}} = (4,3,1,0)^{T}$, and  $\lambda_{2A} = (4,5,4,6)^{T}$.

\subsection{Distribution phase}
\begin{enumerate}
    \item $D$ chooses a random vector $\rho  = (2,5,1,4)^{T}$ and calculate $sh = M\rho = (4,5,6,3)^{T} $.

    \item The dealer $D$ prepares diagonal matrix $\Sigma$ of order $8$ with diagonal elements $sh_{k},~k=1,2,3,4$.
     \begin{equation*}
      \Sigma =   \text{diag}\Big\{4,4,5,5,6,6,3,3\Big\}.
    \end{equation*}
$D$ develops an invertible matrix $Y$ of order $8$ and computes $X = Y^{-1} \Sigma Y$.
\begin{equation*}
Y= \begin{bmatrix} 0 & 0 & 0 & 1 & 0 & 0 & 0 & 0 \\ 0 & 0 & 1 & 0 & 0 & 0 & 0 & 0 \\0 & 1 & 0 & 0 & 0 & 0 & 0 & 0 \\1 & 0 & 0 & 0 & 0 & 0 & 1 & 0 \\ 0 & 0 & 0 & 0 & 0 & 0 & 1 & 0 \\ 0 & 0 & 0 & 0 & 1 & 0 & 0 & 0  \\ 0 & 0 & 1 & 0 & 0 & 1 & 0 & 0  \\ 1 & 0 & 0 & 0 & 0 & 0 & 0 & 1      
\end{bmatrix},
X = Y^{-1} \Sigma Y = 
\begin{bmatrix} 5 & 0 & 0 & 0 & 0 & 0 & -1 & 0 \\ 0 & 5 & 0 & 0 & 0 & 0 & 0 & 0 \\0 & 0 & 4 & 0 & 0 & 0 & 0 & 0 \\0 & 0 & 0 & 4 & 0 & 0 & 1 & 0 \\ 0 & 0 & 0 & 0 & 6 & 0 & 0 & 0 \\ 0 & 0 & -1 & 0 & 0 & 3 & 0 & 0  \\ 0 & 0 & 0 & 0 & 0 & 0 & 6 & 0  \\ -2 & 0 & 0 & 0 & 0 & 0 & 1 & 3      
\end{bmatrix}.
\end{equation*}
The dealer $D$ occurs the linearly independent eigenvectors $(y_{k1},y_{k2})$ corresponding to the eigenvalues $sh_{k}$ of $X$ as secret shares' shadows of participant $P_{k}$. Thus, $D$ transmits these shadows $(y_{k1},y_{k2})$ to the participants $P_{k},~ k=1,2,3,4$.\\
The eigenvectors $(y_{11},y_{12})$ corresponding to the eigenvalue $sh_{1}=4$, are 
\begin{equation*}
   y_{11} = (0,0,1,0,0,-1,0,0)^{T}, y_{12} = (0,0,0,1,0,0,0,0)^{T}.
\end{equation*}
Similarly, the eigenvectors $(y_{21},y_{22})$,$(y_{31},y_{32})$, and $(y_{41},y_{42})$ corresponding to the eigenvalue $sh_{2}=5$,~$sh_{3}=6$, and $sh_{4}=3$, respectively, are 
\begin{equation*}
   y_{21} = (1,0,0,0,0,0,0,-1)^{T}, y_{22} = (0,1,0,0,0,0,0,0)^{T}
\end{equation*}
\begin{equation*}
   y_{31} = (0,0,0,0,1,0,0,0)^{T}, y_{32} = (-1,0,0,0,0,0,1,1)^{T}
\end{equation*}
\begin{equation*}
   y_{41} = (0,0,0,0,0,0,0,1)^{T}, y_{42} = (0,0,0,0,0,1,0,0)^{T}
\end{equation*}

\item $D$ computes the hash values $H_{1}= h(s_{1})$, $H_{2}= h(s_{2}) $,  where $h()$ is a publicly known hash function.

\end{enumerate}

\subsection{Reconstruction phase}
Suppose the participants $A_{1}= \{P_{1},P_{2},P_{3}\}\in\Gamma_{1}$, and $A= \{P_{1},P_{2},P_{3},P_{4}\}\in\Gamma_{2}$ want to retrieve the secrets $s_{1}=2$ and $s_{1}=5$, respectively. 
\subsubsection{Cheating identification phase}

\begin{enumerate}
    \item The secret shares $sh = (sh_{1},sh_{2},sh_{3},sh_{4})^{T} = (4,5,6,3)^{T} $ and the diagonal matrix $\Sigma$ are stored in the Black box.
    \item The participants $P_{k},~ k=1,2,3,4$ provides the secret shares' shadows $(y_{k1},y_{k2})$. The Black box verifies the participants' cheating by the following conditions:
     \begin{itemize}
         \item[$\bullet$] $y_{k1}$ and $y_{k2}$ are linearly independent.
         \item[$\bullet$] $sh_{k} = sh_{k1} = sh_{k2}$.
    \end{itemize}
    If the participants $P_{k},~ k=1,2,3,4$ meet the above two requirements; they will receive their secret shares $sh_{k}$ to recover the secrets $s_{1}=2$ and $s_{2}=5$. Otherwise, they will be identified as cheaters and eliminated.    
\end{enumerate}

\subsubsection{Secret recovery phase}
Suppose that the participants $A_{1}= 
\{P_{1},P_{2},P_{3}\}\in\Gamma_{1}$ want to retrieve the secret $s_{1}$, and $P_{1}$ is a reconstructor. $P_{1}$ generates $3$ single qudit particles and computes $ \ket{\phi_{2}} = \frac{1}{\sqrt{7}} \sum_{v=0}^{6} \ket{v}_{1}\ket{v}_{2}\ket{v}_{3}$. $P_{1}$ transmits $\ket{v}_{2}$ and $\ket{v}_{3}$ to the participants  $P_{2}$ and $P_{3}$, respectively. Now, each participants $P_{1},P_{2}$ and $P_{3}$ applies generalized Pauli operators and get the quantum state $\ket{\phi_{3}}$.
 \begin{align} \nonumber
      \ket{\phi_{3}} &= \frac{1}{\sqrt{7}} \sum_{v=0}^{6} U_{0,2}\ket{v}_{1} \otimes U_{0,1}\ket{v}_{2}\otimes U_{0,6}\ket{v}_{3}\\ \nonumber &
             = \frac{1}{\sqrt{7}} \sum_{v=0}^{6} \omega^{2v}\ket{v}_{1}\omega^{v}\ket{v}_{2}...\omega^{6v}\ket{v}_{t} \\ &
             = \frac{1}{\sqrt{7}} \sum_{v=0}^{6} \omega^{(2+1+6)v} \ket{v}_{1}\ket{v}_{2}\ket{v}_{3}.
   \end{align}
Now, every participant performs the inverse quantum Fourier transform $\mathcal{F}^{-1}$  on their respective particle and then measures the outcome of the $\mathcal{F}^{-1}$ transformation. After performing the measurement, each participant publicly shares their measurement outcome and combines the results. Then, they calculate the secret $s_{1}$ and check the recovered secret by $H_{1} = h(s_{1})$.
\begin{equation}
    \sum_{j=1}^{3}\lambda_{j}sh_{j} \text{mod} 7 = \lambda_{1}sh_{1} + \lambda_{2}sh_{2} +\lambda_{3}sh_{3} = (2+1+6) \text{mod} 7 = 2.
\end{equation}
Similarly, the participants $A= \{P_{1},P_{2},P_{3},P_{3}\}\in\Gamma_{2}$ reconstruct the secret $s_{2}$ and verify the secret by $H_{2} = h(s_{2})$.
\begin{equation}
    \sum_{j=1}^{4}\lambda_{j}sh_{j} \text{mod} 7 = \lambda_{1}sh_{1} + \lambda_{2}sh_{2} +\lambda_{3}sh_{3}+\lambda_{4}sh_{4} = (1+3+6+2) \text{mod} 7 = 5.
\end{equation}

\section{Efficiency analysis in noisy environment}
\begin{figure}[htbp!]
     \centering
     \begin{subfigure}[htbp!]{0.32\textwidth}
         \centering
         \includegraphics[width=\textwidth]{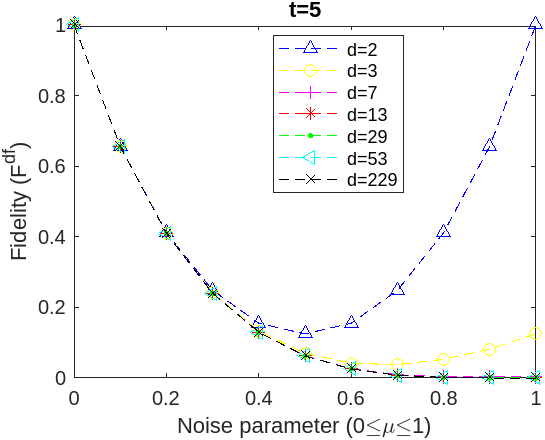}
         \subcaption{Effect of dit-flip noise }
         \label{a}
     \end{subfigure}
     \hfill
     \begin{subfigure}[htbp!]{0.32\textwidth}
         \centering
         \includegraphics[width=\textwidth]{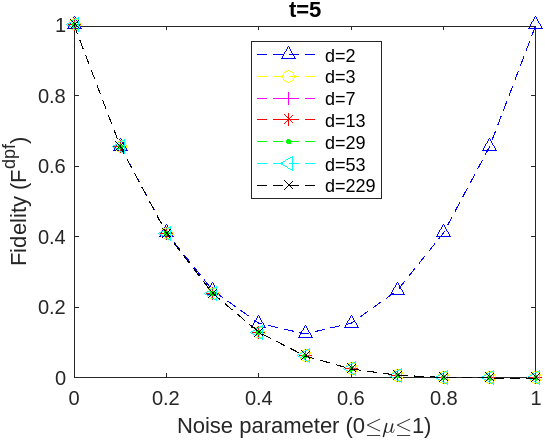}
         \subcaption{Effect of $d$-phase-flip noise}
         \label{b}
     \end{subfigure}
     \hfill
     \begin{subfigure}[htbp!]{0.32\textwidth}
         \centering
         \includegraphics[width=\textwidth]{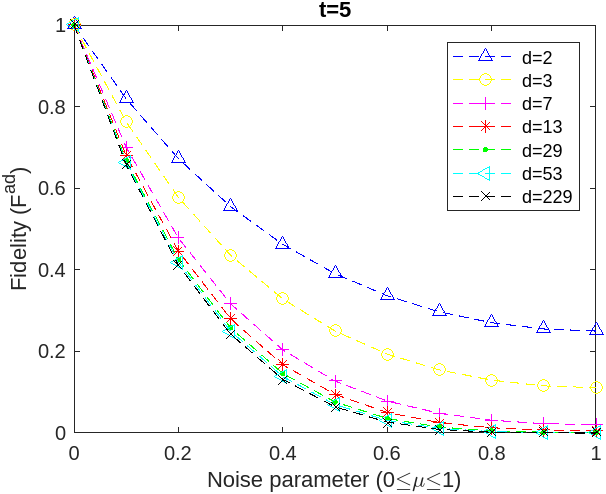}
         \subcaption{Effect of amplitude damping noise}
         \label{c}
     \end{subfigure}
     \newline
      \centering
     \begin{subfigure}[htbp!]{0.32\textwidth}
         \centering
         \includegraphics[width=\textwidth]{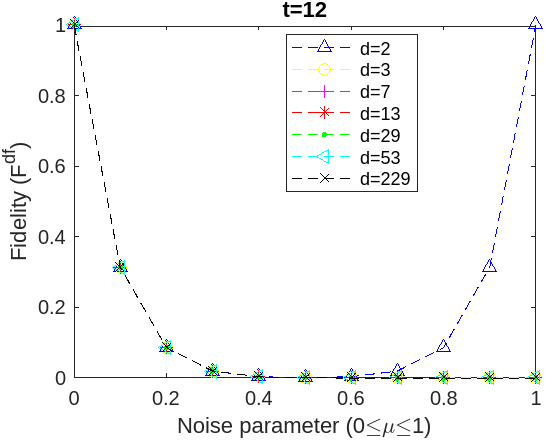}
         \subcaption{Effect of dit-flip noise}
         \label{d}
     \end{subfigure}
     \hfill
     \begin{subfigure}[htbp!]{0.32\textwidth}
         \centering
         \includegraphics[width=\textwidth]{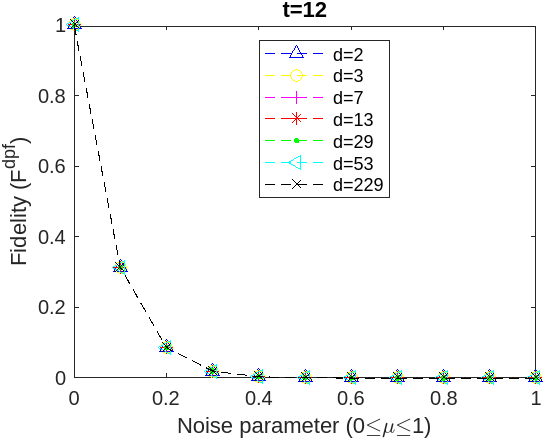}
         \subcaption{Effect of $d$-phase-flip noise}
         \label{e}
     \end{subfigure}
     \hfill
     \begin{subfigure}[htbp!]{0.32\textwidth}
         \centering
         \includegraphics[width=\textwidth]{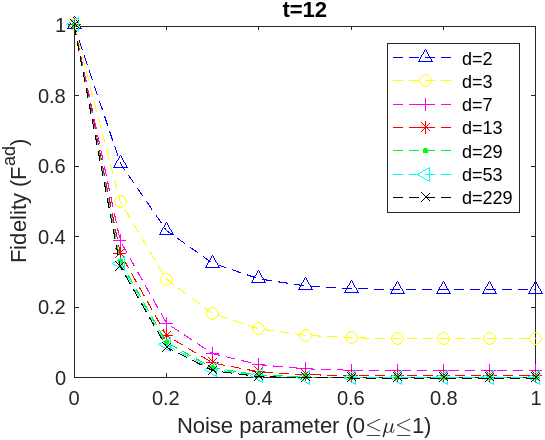}
         \subcaption{Effect of amplitude damping noise}
         \label{f}
     \end{subfigure}
        \caption{The effect of three noises on QMSS by examining the changes in fidelity $F^{r}$ with respect to the noise parameter $\mu$.}
        \label{fig1}
\end{figure}

 In the current advanced quantum technologies, the dealer and participants are expected to create the quantum state accurately. However, when quantum particles are transmitted between the dealer and participants over a quantum channel, channel noise affects QSS protocol execution. Thus, we demonstrate the effectiveness of the proposed scheme in various noise conditions, including $\text{dit-flip (df)}$, $d\text{-phase-flip (dpf)}$, and $\text{amplitude damping (ad)}$.
 \\
 To simplify the study, assume that the participants' local particles are unaffected by channel noise. Only the same kind of noise and similar noise parameters operate on the particles when sent through the quantum channel. In the proposed scheme, the qudits in quantum state $\ket{\phi_{2}}$ are distributed to participants via the quantum channel, and the final quantum state $\ket{\phi_{3}}$ is prepared using local unitary operations. Consequently, the effectiveness of the proposed QMSS protocol relies on the proximity between the final quantum state $\ket{\phi_{3}}$ and the output density matrix $\rho_{out}$.
\\
In proposed protocol, the dealer $D$ splits the secret shadows to the participants $P_{1},P_{2},...,P_{m}$ and assume that the participants of a set $A\in \Gamma_{i}$ retrive the secret $s_{i}$. After the cheating verification of participants, the participant $P_{1}$ prepares the entangled quantum state $\ket{\phi_{2}}$. Therefore, the density matrix of the quantum state $\ket{\phi_{2}}$ is $\rho = \ket{\phi_{2}} \bra{\phi_{2}}$. Thus, $P_{1}$ communicates the states' particles to the participants $P_{j}~(j=2,3,...,t)$. The noise model describing the entire quantum system is presented as follows:
 \begin{equation}
     \rho_{1}^{r} = \epsilon^{r}(\rho)= \sum_{m',n'} (I \otimes E_{m',n'}^{2}\otimes E_{m',n'}^{3} \otimes...\otimes E_{m',n'}^{t}) \rho (I \otimes E_{m',n'}^{2}\otimes E_{m',n'}^{3} \otimes...\otimes E_{m',n'}^{t})^{\dagger}
 \end{equation}
 where $r\in\{df,dpf,ad\}$ for dit-flip, $d$-phase-flip and amplitude damping  noise environments, respectively. After participants $P_{2},P_{3},...,P_{t}$ receive the transmitted particles, the affected density matrices under the dit-flip, $d$-phase-flip, and amplitude damping noise channel can be described as, respectively.
 \begin{equation}
 \begin{aligned}
      \rho_{1}^{df} & = \epsilon^{df}(\rho) =  (I \otimes E_{0,0}^{2}\otimes E_{0,0}^{3} \otimes...\otimes E_{0,0}^{t})\rho (I \otimes E_{0,0}^{2}\otimes E_{0,0}^{3} \otimes...\otimes E_{0,0}^{t})^{\dagger}  \\ &
      + (I \otimes E_{0,1}^{2}\otimes E_{0,1}^{3} \otimes...\otimes E_{0,1}^{t})
      \rho (I \otimes E_{0,1}^{2}\otimes E_{0,1}^{3} \otimes...\otimes E_{0,1}^{t})^{\dagger} +...+   \\ &
      (I \otimes E_{0,d-1}^{2}\otimes E_{0,d-1}^{3} \otimes...\otimes E_{0,d-1}^{t})\rho (I \otimes E_{0,d-1}^{2}\otimes E_{0,d-1}^{3} \otimes...\otimes E_{0,d-1}^{t})^{\dagger} \\ &
      = \frac{1}{d}\Big[ (1-\mu )^{t-1}\Big( \sum_{v=0}^{d-1} \ket{v}_{1} \ket{v}_{2}...\ket{v}_{t}\Big)\Big(\sum_{v=0}^{d-1} \ket{v}_{1} \ket{v}_{2}...\ket{v}_{t}\Big)^{\dagger} + \Big( \frac{\mu}{d-1} \Big)^{t-1} \\ &
      \Big( \sum_{v=0}^{d-1} \ket{v}_{1}\ket{v+1}_{2}...\ket{v+1}_{t} \Big) 
      \Big(\sum_{v=0}^{d-1}\ket{v}_{1} \ket{v+1}_{1}...\ket{v+1}_{t} \Big)^{\dagger} 
      +...+ \Big( \frac{\mu}{d-1} \Big)^{t-1}  \\ &
      \Big( \sum_{v=0}^{d-1} \ket{v}_{1} \ket{v+d-1}_{2}...\ket{v+d-1}_{t} \Big) 
      \Big( \sum_{v=0}^{d-1} \ket{v}_{1} \ket{v+d-1}_{2}...\ket{v+1}_{t} \Big)^{\dagger}
      \Big]
  \end{aligned}
   \end{equation}
   
 \begin{align} \nonumber
    \rho_{1}^{dpf} & = \epsilon^{dpf}(\rho) =  (I \otimes E_{0,0}^{2}\otimes E_{0,0}^{3} \otimes...\otimes E_{0,0}^{t})\rho (I \otimes E_{0,0}^{2}\otimes E_{0,0}^{3} \otimes...\otimes E_{0,0}^{t})^{\dagger} \\ \nonumber &
      + (I \otimes E_{1,0}^{2}\otimes E_{1,0}^{3} \otimes...\otimes E_{1,0}^{t}) 
      \rho (I \otimes E_{1,0}^{2}\otimes E_{1,0}^{3} \otimes...\otimes E_{1,0}^{t})^{\dagger} +...+ \\ \nonumber & 
      (I \otimes E_{d-1,0}^{2}\otimes E_{d-1,0}^{3} \otimes...\otimes E_{d-1,0}^{t})\rho (I \otimes E_{d-1,0}^{2}\otimes E_{d-1,0}^{3} \otimes...\otimes E_{d-1,0}^{t})^{\dagger} \\ \nonumber &  
    = \frac{1}{d}\Big[ (1-\mu )^{t-1}\Big(\sum_{v=0}^{d-1} \ket{v}_{1} \ket{v}_{2}...\ket{v}_{t}\Big)\Big(\sum_{v=0}^{d-1} \ket{v}_{1} \ket{v}_{2}...\ket{v}_{t}\Big)^{\dagger}+ \Big( \frac{\mu}{d-1} \Big)^{t-1} \\ \nonumber &
    \Big( \sum_{v=0}^{d-1} \omega^{(t-1)v} \ket{v}_{1} \ket{v}_{2}...\ket{v}_{t}\Big) 
    \Big( \sum_{v=0}^{d-1} \omega^{(t-1)v} \ket{v}_{1} \ket{v}_{2}...\ket{v}_{t}\Big)^{\dagger} 
    + \Big( \frac{\mu}{d-1} \Big)^{t-1} \\ \nonumber &
    \Big( \sum_{v=0}^{d-1} \omega^{2(t-1)v} \ket{v}_{1} \ket{v}_{2}...\ket{v}_{t}\Big) \Big( \sum_{v=0}^{d-1} \omega^{2(t-1)v} \ket{v}_{1} \ket{v}_{2}...\ket{v}_{t}\Big)^{\dagger} 
    +...+ \Big( \frac{\mu}{d-1} \Big)^{t-1} \\ &
    \Big( \sum_{v=0}^{d-1} \omega^{(d-1)(t-1)v} \ket{v}_{1} \ket{v}_{2}...\ket{v}_{t}\Big)\Big( \sum_{v=0}^{d-1} \omega^{(d-1)(t-1)v} \ket{v}_{1} \ket{v}_{2}...\ket{v}_{t}\Big)^{\dagger}  \Big]
 \end{align}
 \begin{align} \nonumber
     \rho_{1}^{ad} & = \epsilon^{ad}(\rho) =  (I \otimes E_{0}^{2}\otimes E_{0}^{3} \otimes...\otimes E_{0}^{t})\rho (I \otimes E_{0}^{2}\otimes E_{0}^{3} \otimes...\otimes E_{0}^{t})^{\dagger}
      + (I \otimes E_{1}^{2}\otimes E_{1}^{3} \\ \nonumber &
      \otimes...\otimes E_{1}^{t})\rho (I \otimes E_{1}^{2}\otimes E_{1}^{3} 
      \otimes...\otimes E_{1}^{t})^{\dagger} +...+  
      (I \otimes E_{d-1}^{2}\otimes E_{d-1}^{3} \otimes...\otimes E_{d-1}^{t})\rho  \\  \nonumber &
      (I \otimes E_{d-1}^{2}\otimes E_{d-1}^{3} \otimes...\otimes E_{d-1}^{t})^{\dagger}  \\ \nonumber &
      = \frac{1}{d}\Big(\ket{0}_{1}\ket{0}_{2}...\ket{0}_{t} + (1-\mu)^{\frac{t-1}{2}} \sum_{v=1}^{d-1} \ket{v}_{1} \ket{v}_{2}...\ket{v}_{t}\Big) \Big(\ket{0}_{1}\ket{0}_{2}...\ket{0}_{t}
      + (1-\mu)^{\frac{t-1}{2}} \\ & 
      \sum_{v=1}^{d-1} \ket{v}_{1} \ket{v}_{2}...\ket{v}_{t}\Big)^{\dagger}  
      + \frac{\mu^{t-1}}{d} \sum_{v=1}^{d-1} (\ket{v}_{1}\ket{0}_{2}...\ket{0}_{t})(\ket{v}_{1}\ket{0}_{2}...\ket{0}_{t})^{\dagger}
 \end{align}
Now, the participants $P_{j},~j=1,2,...,t$ implement the generalized Pauli operator $U_{0,\lambda_{j}sh_{j}}$ on their particles and the resultant density matrix defined by:
\begin{align}
  \rho_{out}^{r} =  (U_{0,\lambda_{1}sh_{1}} \otimes U_{0,\lambda_{2}sh_{2}}\otimes...\otimes U_{0,\lambda_{t}sh_{t}}) \rho_{1}^{r} (U_{0,\lambda_{1}sh_{1}} \otimes U_{0,\lambda_{2}sh_{2}}\otimes...\otimes U_{0,\lambda_{t}sh_{t}})^{\dagger} 
\end{align}
In an ideal situation with no noise in the quantum channel, all participants can generate the quantum state $\ket{\phi_{3}}$. The effectiveness of the proposed QMSS scheme under various noise conditions can be assessed by measuring the fidelity between the output density matrix $\rho_{out}^{r}$ and the quantum state $\ket{\phi_{3}}$. The fidelity in the various noises can be characterized as follows. 
\begin{align}
   & F^{df} = (1-\mu)^{t-1} \\ &
   F^{dpf} = \begin{cases}
        (1-\mu)^{t-1}+\frac{\mu^{t-1}}{(d-1)^{t-2}}, & (t-1) \overset{d}{\equiv}0; \\ 
        (1-\mu)^{t-1}, & (t-1) \overset{d}{\not\equiv}0
        \end{cases}      ~~ \text{(Using remark 3)}   \\ &
    F^{ad} = \frac{1}{d^2}(1+(1-\mu)^{\frac{t-1}{2}}(d-1))^{2}
\end{align}
To illustrate the fidelity in different noise environments, we examine the performance by taking $t=5,12$ participants with dimensions $d=2,3,7,13,29,53,229$. The fidelity of three distinct noise models is determined using MATLAB in Fig.(\ref{fig1}). The influence of dit-flip noise is shown in Figs. (\ref{a}) and (\ref{d}) by a graphical depiction of the change in fidelity compared to the noise parameter $\mu$. The graph shows that when $t=5$, fidelity decreases with increasing noise parameter $\mu$ and gets to zero as $\mu\in[0.8,1]$. The fidelity $F^{df}$ does not vary as the dimension $d$ increases since it is independent of the quantum system's dimension. Following that, when $\mu$ grows, the fidelity $F^{df}$ reduces rapidly for the larger number of participants $t$ (larger number of qudit particles). In the instance of $t=12$, $F^{df}$ reaches zero when $\mu\in[0.4,1]$.

The graphical depiction Figs.(\ref{b}) and (\ref{e}) of $d$-phase-flip fidelity $F^{dpf}$ and noise parameter $\mu$ show that for $t=5$ and $d=2$ dimensions, the fidelity $F^{dpf}$ decreases as the noise parameter $\mu\in[0,0.5]$ increases. Following that, $F^{dpf}$ begins to increase and reaches $1$ as $\mu\in[0.5,1]$. However, at larger dimensions $d=3,7,13,29,53,229$, the fidelity $F^{dpf}$ declines as $\mu$ increases and reaches zero as $\mu\in[0.65,1]$.  For $t=12$ participants, $F^{dpf}$ immediately falls to zero for $\mu\in[0.4,1]$.  
\\
The graph in Fig.(\ref{c}) of amplitude damping fidelity $F^{ad}$ and noise parameter $\mu$ demonstrates that for $t=5$ and $d=2,3,7$, the fidelity $F^{ad}$ falls as the noise parameter $\mu$ increases. Whereas for higher dimensions $d=13,29,53,229$, $F^{ad}$ rapidly decrease and becomes zero as $\mu\in[0.7,1]$. In the instance of $t=12$, $F^{ad}$ decreases rapidly as $\mu$ increases for $d=2,3,7$ and approaches to zero as $\mu\in[0.7,1]$ for $d=13,29,53,229$ given in Fig.(\ref{f}).
\\
In three distinct noise environments, the effectiveness of the proposed QMSS protocol decreases as the noise parameter $\mu$ increases. Nevertheless, within the range of noise parameter $\mu\in[0,0.4]$, the proposed scheme demonstrates superior efficiency in the presence of amplitude-damping noise compared to the dit-flip and $d$-phase-flip noise channels. For $t=5$ and $d=2$, the proposed scheme exhibits greater efficiency in the presence of $d$-phase-flip noise, precisely when $\mu\in[0.8,1]$, compared to other noise channels. The fidelity of dit-flip and $d$-phase-flip noises are correlated in some instances.

\section{Comparisons}
This section compares the proposed QMSS protocol with several other similar existing $d$-dimensional QSS protocols\cite{qin2018multi,mashhadi2019general,sutradhar2021enhanced,li2021general,mashhadi2022verifiable,yan2022cheating}. Qin et al.\cite{qin2018multi} developed a multi-dimensional QSS protocol using SUM operator and quantum Fourier transform for encoding and decoding the qudit state as a secret. The dealer allocates the particles among $m$ participants, with $m-1$ participants performing measurements on their particles, while the last participant applies a unitary operation to his particle depending on the measurement outcomes. However, this $(m,m)$ threshold scheme is vulnerable to participant attacks such as forgery and collusion and cannot detect dishonest behavior of the participants, making it less flexible and less secure. Mashhadi\cite{mashhadi2019general} presented a QSS scheme utilizing the quantum Fourier transform and general access structure to share a classical secret. The scheme requires one trusted player in each authorized set to reconstruct the secret. While the scheme demonstrates resilience against various attacks and the ability to detect cheating, it cannot distinguish malicious participants. Sutradhar and Om\cite{sutradhar2021enhanced} suggested an enhancement to the QSS scheme\cite{song2017t} by introducing a $(t,m)$-threshold QSS scheme. Nonetheless, the scheme is restricted to a $(t,m)$-threshold and cannot identify dishonest participants.\\
 Based on the two qudit generalized Bell states, a general QSS technique was introduced by Li et al.\cite{li2021general}. In this technique, participants reconstruct the secret by operating a generalized Pauli operator. Yan et al.\cite{yan2022cheating} developed a $(t,m)$-threshold QSS protocol with cheat-identification of the participants. The dealer provides two identical quantum states, one signed for secret sharing and the other for identifying cheating. The participants apply unitary transformations on two quantum states and verify the cheating of successive participants by quantum digital signature mechanism. Nonetheless, this protocol presents implementation challenges, and its practical application may be limited.
Furthermore, in the schemes\cite{qin2018multi,mashhadi2019general,sutradhar2021enhanced,li2021general,yan2022cheating}, the dealer is only capable of sharing a single classical secret with participants. In contrast, our proposed protocol enables the sharing of multiple secrets simultaneously to different subsets of participants. Mashhadi\cite{mashhadi2022verifiable} presented a QSS scheme that employs a single qudit state and unitary operations to share multiple classical secrets with multiple access structures. He examines the internal eavesdropping using a memoryless qudit quantum channel and the weak locking for the erasure channel. Thus, this scheme cannot detect the dishonest participant. Furthermore, none of the above QSS schemes have been observed in noise environments.\\
In contrast, we propose a cheat-detection QSS protocol to share multiple classical secrets with different subsets of participants. The Black box's cheat-detection technology can recognize and identify each participant's deceptive behavior. The scheme can endure several typical attacks, such as forgery and collusion attacks, making it more secure. Furthermore, we emphasize the efficiency of the scheme in various noise environments. Consequently, the proposed scheme features a robust cheat-detection technique assures the honesty of participants, thereby enhancing overall security and also demonstrating its effectiveness in noisy environments. Table \ref{tab.1} compares our proposed QMSS protocol and other recently developed QSS schemes.
\begin{landscape}
\begin{table}[htbp!]
		\caption{Comparison of schemes}
		\label{tab.1}
		\label{aggiungi}\centering \footnotesize
            
		\begin{tabular}{lp{50pt}p{50pt}p{50pt}p{50pt}p{50pt}p{50pt}p{50pt}p{15pt}}
		\midrule
        
			Parameters  &  Qin\cite{qin2018multi} &  Mashhadi\cite{mashhadi2019general} &  Sutradhar\cite{sutradhar2021enhanced} & Li\cite{li2021general} & Yan\cite{yan2022cheating} & Mashhadi\cite{mashhadi2022verifiable} & Proposed \\ \toprule
		Structure &  $(m,m)$  & General   &  $(t,m)$  & General   &  $(t,m)$  & General & General    \\                   
                      &  threshold &     & threshold &    & threshold &   &    \\
		
            Dimension of the space &  $d$ & $d$ & $d$ & $d$ & $d$ & $d$  & $d$                                     \\
            
		Secret type   & Quantum state  &  Classical info. & Classical info. &  Classical info. and & Classical info.   & Classical info.   & Classical info. \\
                          &            &          &         &     Quantum (single qubit)        &             &   &  \\  
     Number of secrets    & one & one & one & one & one & $n$ & $n$ &   \\        
                
         Eavesdropping check  &  Decoy  &   Decoy &  $-$  &  $-$  &  Quantum digital    &  Decoy  &  Black box                \\
                             & particles & particles &  &     &  signature    &  particles &             \\
                             
         Verification of secret & $-$  & Hash $\text{function}$  & Hash $\text{function}$ & Hash $\text{function}$ & Hash $\text{function}$  & Hash $\text{function}$ & Hash $\text{function}$   \\
         
          Entanglement used  &   No   &   No  &   No &   Yes  & Yes  &   No &      No   \\
          
          Quantum operation  &  QFT, Pauli, ${\text{QFT}}^{-1}$ & QFT, Pauli &  QFT, Pauli, ${\text{QFT}}^{-1}$ & Pauli & Unitary operators & Pauli & QFT, Pauli, ${\text{QFT}}^{-1}$ \\
               
         Cheating identifiable  &   No   &  No  &    No &    No &    Yes   & No    & Yes \\  
         
        Total no. of operations    &  $t(t+1)+m+2$  & $2t$ & $t$ & $t$ & $t+1$  & $t+1$ & $t$ \\

         Measurement operations &   $m$   &   $t$   & $t-1$   & $t$  & $t+2$  & $1$ & $t-1$ \\   
         Efficiency analysis      & No & No & No & No & No & No & Yes \\
         under noise environments &    &    &    &    &    &    & \\
              \midrule
		\end{tabular}
	\end{table}
\end{landscape}

\section{Conclusions}
This study presents a $d$-dimensional QMSS protocol with cheat identification using multi-access structures and a monotone span program. The dealer distributes multiple classical secrets to participants, and authorized sets of participants retrieve them by utilizing QFT and unitary operators. The deception verification mechanism in the Black box can identify each participant's dishonesty. The security evaluation demonstrates that the proposed approach is resistant to various attacks, intercept resend, entangle measure, and participant attacks, including forgery and collusion. Furthermore, the proposed protocol's efficiency is evaluated using quantum fidelity in several noise models: dit-flip, $d$-phase-flip, and amplitude-damping. Compared to existing QSS protocols, the proposed scheme has several characteristics: to detect dishonest participants, no need for entanglement measurement, more straightforward implementation, greater efficiency, and practicality under real-world conditions. 



\paragraph{Acknowledgement}\small
The first author, supported by grant number 09/143(0951)/2019-EMR-I, expresses gratitude to the Council of Scientific and Industrial Research (CSIR), India, for their financial assistance in conducting this work. This research is also supported by SERB core grant number CRG/2020/002040.
\\

\noindent
\small{\textbf{Data availability}}\\
Data sharing is not applicable to this article as no datasets were generated or analyzed during the current study.\\

\noindent
\textbf{Declaration of competing interests}\\
The authors have no competing interests to declare that are relevant to the content of this article.


 \end{document}